# On the origin of blueshifts in organic polariton condensates

Timur Yagafarov[1], Denis Sannikov[1,2], Anton Zasedatelev[1,3*], Kyriacos Georgiou[4], Anton Baranikov[1], Oleksandr Kyriienko[5,6], Ivan Shelykh[6,7], Lizhi Gai[8], Zhen Shen[8], David G. Lidzey[4], and Pavlos G. Lagoudakis[1,3*]

*1. Center of Photonics and Quantum Materials, Skolkovo Institute of Science and Technology, Moscow 121205, Russian Federation.*

*2. Lebedev Physical Institute of the Russian Academy of Sciences, Moscow 119333, Russian Federation.*

*3. Department of Physics and Astronomy, University of Southampton, Southampton SO17 1BJ, United Kingdom.*

*4. Department of Physics and Astronomy, University of Sheffield, Sheffield S3 7RH, United Kingdom.*

*5. NORDITA, KTH Royal Institute of Technology and Stockholm University, Stockholm SE-106 91, Sweden.*

*6. ITMO University, Saint Petersburg 197101, Russian Federation.*

*7. School of Engineering and Natural Sciences, University of Iceland, Reykjavik IS-107, Iceland.*

*8. State Key Laboratory of Coordination and Chemistry, School of Chemistry and Chemical Engineering, Nanjing University, Nanjing 210046, China*

[*] Authors for correspondence: A.Zasedatelev@skoltech.ru; pavlos.lagoudakis@soton.ac.uk

**Abstract**

We report on the origin of energy-shifts in organic polariton condensates. The localised nature of Frenkel excitons in molecular semiconductors precludes interparticle Coulomb exchange interactions – the latter being the dominant mechanism for blueshifts in inorganic semiconductor microcavities that bear Wannier-Mott excitons. We examine the contribution of optically induced change of the intracavity non-linear refractive index, gain induced frequency-pulling and quenching of the Rabi splitting, as well as the role of polariton-exciton and polariton-polariton scattering in the energy-shift of the polariton mode at condensation threshold in strongly coupled molecular dye microcavities. We conclude that blueshifts in organic polariton condensates arise from the interplay of the saturation of molecular optical transitions and intermolecular energy migration. Our model predicts the commonly observed step-wise increase of both the emission energy and degree of linear polarisation at polariton condensation threshold.

Electronic excitations in organic semiconductors can be described in terms of Frenkel excitons in which strongly-bound electron-hole pairs are localized on a single molecule. Frenkel excitons typically have a binding energy of the order of 0.5 - 1 eV that allows them to survive at ambient temperatures. Following the first observation of strong coupling in organic microcavities (MCs) [1], the rapid development of organic polaritonics has culminated in the observation of polariton condensates at room temperature. Emission wavelengths spanning the visible spectrum have been demonstrated through the selection of appropriate molecular materials [2–7]. Furthermore, polariton condensation in organic microcavities has led to the demonstration of polariton transistors operating at ambient conditions having record-high optical gain (~10dB/µm) [8]. The coherent nature of organic polariton condensates also results in superfluid propagation [9], suggesting further applications in polariton circuits.

Despite the rapid progress of organic polaritonics and their potential for applications, the mechanisms underlying polariton nonlinearities remain poorly understood. In organic semiconductors, the localization of Frenkel excitons on a single molecule dramatically weakens Coulomb exchange interactions and interparticle scattering. The blueshift of a polariton mode is considered a key manifestation of polariton interactions in inorganic microcavities [10,11]. However in organic polariton systems, the origin of a stepwise energy-shift observed at condensation thresholds remains unclear, despite its omnipresence across a diverse range of organic materials [3–8].

In this letter, we explore the origin of blueshifts in organic polariton condensates. We examine the relative contribution of a number of processes, including optically induced changes of the intracavity non-linear refractive index, gain induced frequency-pulling, polariton-exciton and polariton-polariton scattering, as well as the quenching of the Rabi splitting due to the saturation of molecular optical transitions. Through a quantitative analysis, we conclude that the blueshifts in organic polariton condensates and the step-wise energy increase, observed at threshold [3–8], result from an interplay between stimulated relaxation to the polariton ground-state and intermolecular energy transfer. The latter process results in a depolarization of the emission with respect to the polarization of the excitation beam. Our interpretation is qualitatively and quantitatively corroborated by a concomitant step-wise increase of the degree of linear polarization of the emission at condensation threshold [3,4].

The organic microcavities studied here, consisted of a ~$\lambda/2$ spin-cast thin film of BODIPY-G1 dye uniformly dispersed in a polystyrene matrix that was positioned between two distributed Bragg reflectors (DBRs) consisting of 8 and 10 pairs of $SiO_2/Nb_2O_5$ placed on the top and bottom of the structure respectively. For more information about sample fabrication, see Section I in Supplemental Materials (SM). We have found that the spin-casting process used to deposit the organic film results in a gradual increase of film thickness towards the edges of the substrate (bottom mirror). We use this non-uniformity to access a broad range of exciton-photon detuning ($\delta$).

We measure angle-resolved reflectivity of a typical microcavity as shown in Fig. 1(a). Here both upper and lower polariton branches can be observed as local minima in the broad DBR reflectivity stop-band that are split around the BODIPY-G1 peak absorption wavelength at 507 nm. For further experimental details, see Section II in SM. We plot the energy of these modes as a function of angle, creating a dispersion plot as shown in Fig. 1(b) (red-squares). This data is superimposed over a false-color plot of polariton photoluminescence intensity obtained under non-resonant excitation at 400 nm in the linear excitation regime. We fit the upper and lower polariton branches in Fig. 1(b) using a two coupled-oscillator model [12,13], and obtain a vacuum Rabi splitting ($\hbar\Omega_0$) of ~116 meV and an exciton-photon detuning of -160 meV (further details are given in Section III of the SM).

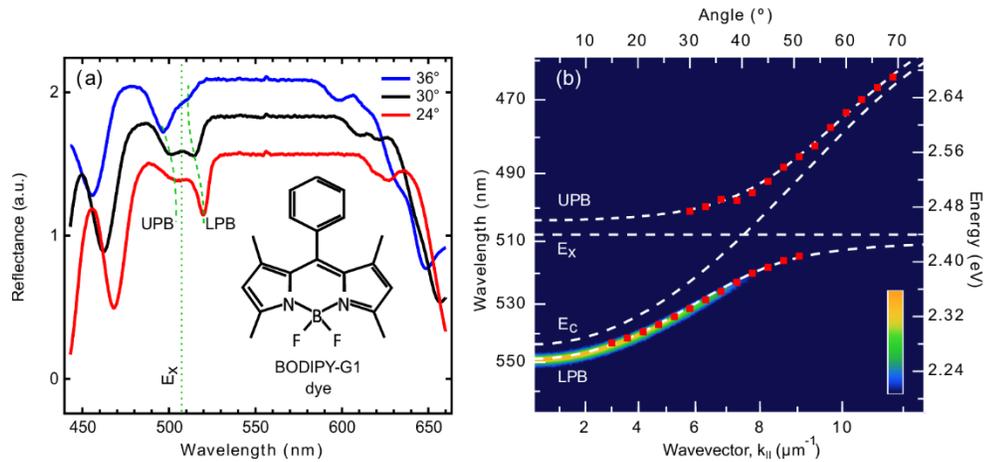

FIG. 1. Strong light-matter interaction in dye-filled microcavities. (a) Angle-dependent reflectivity spectra of the microcavity recorded at 36° (blue), 30° (black) and 24° (red) degrees exhibit clear anti-crossing at the exciton resonance energy $E_x$ (green dotted line) and indicate formation of lower (LPB, green dashed line) and upper (UPB, green dashed line) exciton-polaritons branches. Polariton dispersion relation in (b) is plotted by combing the data of photoluminescence imaging acquired in a Fourier space (rainbow color density plot in a log scale) with the polariton states extracted from angle-dependent reflectivity measurements (red squares). Fits to the LPB and UPB, together with the cavity mode $E_c$ and energy of exciton resonance $E_x$ are shown as white dashed curves. The chemical structure of BODIPY-G1 dye is shown as an inset in (a).

Recent studies have shown that a number of the BODIPY family of molecular dyes undergo polariton condensation/lasing following non-resonant optical excitation [6,7]. To demonstrate condensation using BODIPY-G1, we record the dispersion of polariton photoluminescence emission as a function of excitation density using single excitation pulses in a transmission configuration (see Section II in SM). The excitation laser used provides 2 ps pulses at 400 nm having a horizontal polarization. Fig. 2(a) shows the time-integrated polariton photoluminescence distributed across the lower polariton branch below condensation-threshold. Fig. 2(b) shows microcavity emission above-threshold, where it can be seen that polariton photoluminescence collapses to the bottom of the lower polariton branch. In Fig. 2(c) we plot the photoluminescence intensity and the full width at half maximum of the emission linewidth (right axis, in red) integrated over ±1° (±0.2 μm$^{-1}$) around normal incidence versus excitation density. The corresponding energy-shift and the degree of linear polarization of the emission spectrum are also shown in Figs. 2(d) and (e), respectively. At an excitation density of ~6 mJ/cm$^2$ (120 μJ/cm$^2$ of absorbed pump fluence), we observe a rapid increase of the photoluminescence intensity, a concomitant linewidth narrowing from 1.6 nm to 0.25 nm and a step-wise increase of the degree of linear polarization and a step-wise blueshift of the emission spectrum. Such blueshifts of the polariton emission wavelength occurring around a lasing threshold are considered a hallmark of polariton condensation. In inorganic semiconductor microcavities, such energy-shifts originate from the repulsive interparticle Coulomb exchange interactions between Wannier-Mott excitons [14]. However, such interactions are in principle precluded in polaritons created using molecular semiconductors as a result of the highly localized nature of Frenkel excitons [15,16].

To explore the mechanism behind such blueshifts we examine the contribution of various nonlinear optical phenomena, omnipresent both in the strong- and weak-coupling regimes; namely that of an optically induced change of the intracavity non-linear refractive index and of gain induced frequency-pulling.

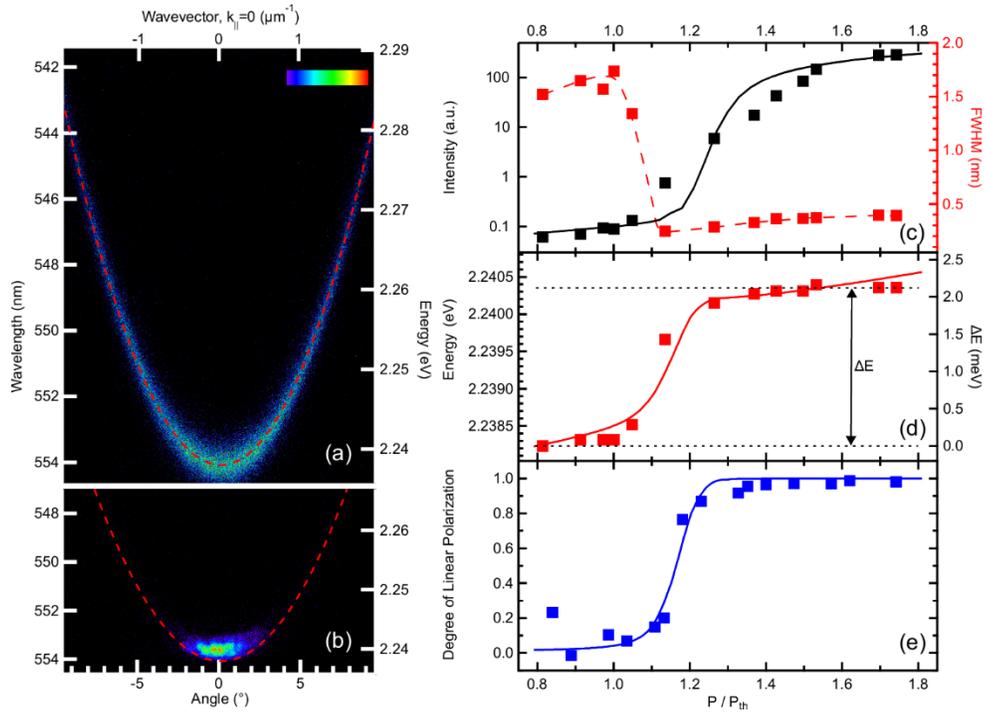

FIG. 2. Non-equilibrium polariton condensation. Normalized $E,k$ polariton-population (photoluminescence) images recorded using Fourier-space imaging, below, at $0.8P_{th}$, (a), and above condensation threshold, at $1.4P_{th}$, (b). Dashed red curves show lower polariton branch dispersion in the linear regime. (c) Photoluminescence intensity at $k_{\parallel} \sim 0$, integrated in the range over $\pm 0.2\ \mu m^{-1}$ (black squares) and full width at half maximum (FWHM, red squares) and (d) energy of the ground polariton state versus pump-power. The superlinear increase and 10-fold line narrowing together with high-energy shift of the polariton ground state observed above threshold are well-known hallmarks of non-equilibrium polariton condensation. (e) Degree of linear polarization (blue squares) as a function of pump power, indicating that the condensate inherits the polarization of the pump beam. Solid curves in (c,d,e) represent numerical simulations of time-integrated photoluminescence, the ground state energy and degree of linear polarization as functions of pump power, respectively. The red dashed curve in (c) is a guide for the eye.

A step-wise increase in the intensity of the electric field inside the cavity at condensation threshold (determined at the lower polariton mode wavelength) could potentially shift the resonance through a change in the nonlinear refractive index of the intracavity material. To examine the contribution of the E-field induced difference in the refractive index, we measure the optical nonlinearities of the bare intracavity medium using both a closed- and open-aperture Z-scan technique [17] (see details in Section II in SM). Fig. 3(a) shows an open-aperture Z-scan transmission recorded at two pulse energies, probing the imaginary part of the nonlinear susceptibility. We find that at the lower incident pulse energy of 9.5 nJ, we do not observe any nonlinear change of the absorption. We note that at the foci, the 9.5 nJ excitation pulse induces an electric-field intensity of 17 GW/cm$^2$; a value that is approximately an order of magnitude greater than the electric-field intensity at the anti-node within the microcavity at condensation threshold (~2 GW/cm$^2$) (calculation shown in Section IV in the SM). At the considerable higher intensity of 779 GW/cm$^2$ at the foci of the beam, we observe an optical nonlinearity in the form of reverse saturable absorption.

The closed-aperture Z-scan measurements presented in Fig. 3(b) provides a measure of the real part of the nonlinear optical susceptibility. Here, we do not detect any change in the nonlinear refractive index when pumping at the lower incident pulse energy. However at the higher incident intensity of 779 GW/cm$^2$, we find that the material exhibits a weak self-focusing effect from which we determine a positive nonlinear refractive index ($n_2$) of ~$1.89 \cdot 10^{-14}$ cm$^2$/W (calculation shown in Section IV in SM). We note that even if such high electromagnetic fields could be generated within a microcavity, the positive value of $n_2$ that we determine would induce a redshift. We conclude therefore that an optically induced change of the intracavity nonlinear refractive index is not responsible for the blueshift observed at condensation threshold.

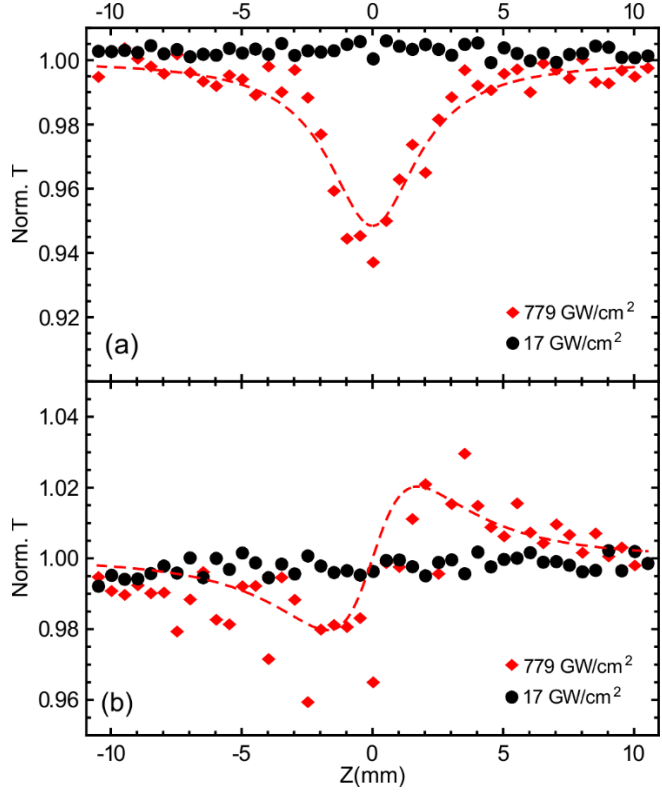

FIG. 3. Nonlinear optical susceptibility of the neat film. (a) Open aperture and (b) closed aperture Z-scan data measured at two different incident pump energies 9.5 nJ (black circles) and 438 nJ (red rhombs). The fitting curves (red dashed) in (a) and (b) correspond to $Im[\chi^{(3)}] = 1.71 \times 10^{-20}\ m^2/V^2$ and $Re[\chi^{(3)}] = 2.17 \times 10^{-20}\ m^2/V^2$, respectively.

We now consider whether gain induced frequency-pulling could be responsible for the blueshift at condensation threshold. This mechanism is expected to be particularly important in negatively detuned microcavities in which the polariton mode has a large photonic fraction. We characterise the spectral distribution of the optical gain by recording amplified spontaneous emission (ASE) from a control (non-cavity) film of BODIPY-G1 dye molecules dispersed in a polystyrene matrix (see experimental methods in Section II in SM). A typical ASE spectrum is plotted in Fig. 4(a) [red line], where it can be seen that the emission (corresponding to the peak of optical gain) is peaked around 2.272 eV (545.8 nm).

We explore the extent to which gain induced frequency-pulling affects the condensate's blueshift by tuning the frequency of the lower polariton branch across the optical gain spectrum. Such tunability in the lower polariton branch wavelength is possible through the variation in the thickness of the intracavity film across the sample. This effect allows us to explore polariton condensation over a broad range of exciton-photon detuning conditions. Fig. 4(b) shows the measured energy-shift for ~400 single-shot measurements of polariton condensation at a wide range of different polariton ground state energies. For each measurement, the energy-shift is defined by comparing the energy of the emission below and above threshold. Here, we avoid averaging over the intensity fluctuations of the laser by utilising a single-shot dispersion imaging technique. In Fig. 4(a), we superimpose the measured blueshift using the Sturge binning rule with the amplified spontaneous emission spectrum (for details see Section III in SM). It is apparent that at condensation threshold, the recorded energy-shifts are always positive and thus we conclude that the blueshift is not induced by gain frequency pulling.

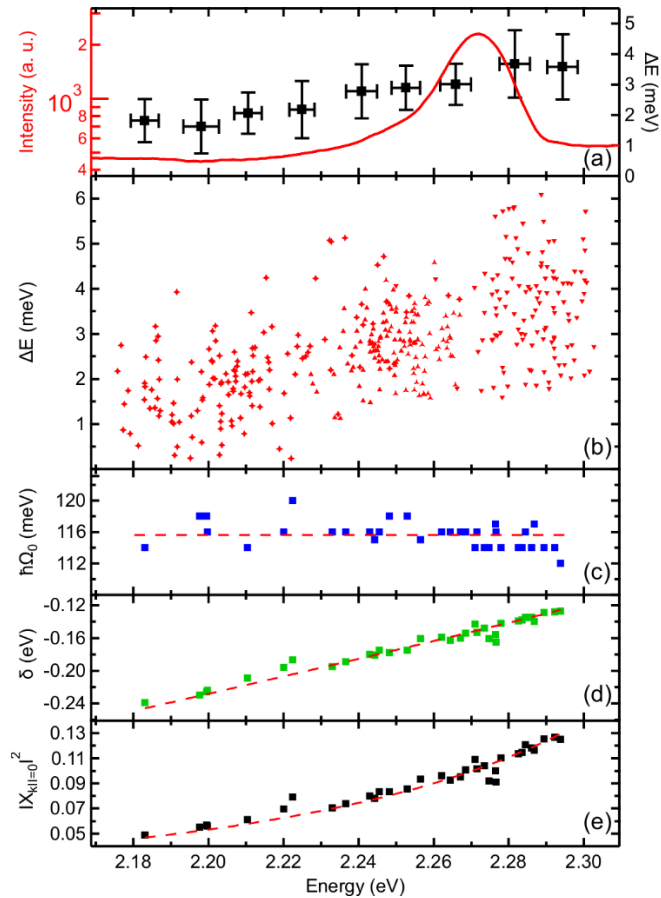

FIG. 4. Blueshift and optical gain spectra. (a) The amplified spontaneous emission spectrum and the blueshift, $\Delta E$, extracted from the binning of scattering plot (b) that shows single-shot blueshift realizations across the whole sample area. (c) Vacuum Rabi splitting $\hbar\Omega_0$ and (d) detuning $\delta$ extracted from the multiple angle-dependent reflectivity measurements carried out across the sample area. The Rabi splitting remains constant across the whole sample area with an average value of (116 ± 1.5) meV (red dashed curve). (e) The exciton fraction $|X_{k_\parallel=0}|^2$ of the polariton wave function at $k_\parallel = 0$, calculated from $\hbar\Omega_0$ and $\delta$ shown above in (c) and (d), respectively. Red dashed curves in (d) and (e) correspond to analytical fit functions for $\delta$ and $|X_{k_\parallel=0}|^2$, respectively.

We now investigate the possible contribution of polariton-exciton and pair-polariton scattering in determining the observed blueshift at condensation threshold. In semiconductor microcavities containing Wannier-Mott excitons, the experimentally observed energy-shifts ($\Delta E$) are attributed to a combination of pair-polariton ($g_{p-p}N_p$) and polariton-exciton ($g_{p-x}N_x$) interaction terms. This is summarized by the following equation [11,18]

$$\Delta E = g_{p-p} \cdot N_p + g_{p-x} \cdot N_x \qquad (1)$$

where the pair-polariton scattering interaction constant can be related to the exciton-exciton scattering constant ($g_{x-x}$) using $g_{p-p} = g_{p-x} \cdot |X|^2 = g_{x-x} \cdot |X|^4$, where $X$ is the amplitude of the exciton-fraction that is mixed into a polariton state, and $N_p$ and $N_X$ are the polariton and exciton reservoir densities, respectively. Since the occupancy of polaritons at condensation threshold does not depend on the exciton fraction, the measured dependence of the energy-shift vs the square of the amplitude of the exciton fraction ($|X|^2$) should reveal whether the blueshift is dominated by pair-polariton or polariton-exciton interactions.

To determine the dependence of the measured blueshift shown in Fig. 4(b) vs the exciton fraction, we need first to describe the dependence of the experimentally-measured emission frequency of the polariton-state on its exciton fraction. The latter depends on the exciton-photon detuning ($\delta$) and vacuum Rabi splitting through $|X_{k_\parallel=0}|^2 = \frac{1}{2}\left(1 + \frac{\delta}{\sqrt{\delta^2+(\hbar\Omega_0)^2}}\right)$. To avoid any excitation density dependent energy-shifts of the lower polariton branch, we perform white light, angle-resolved reflectivity measurements across the available detuning range. We fit the linear polariton dispersions by varying the vacuum Rabi splitting and the exciton-photon detuning, whilst keeping the exciton energy and the effective refractive index of the intracavity layer constant (see Section

III in SM). Figs. 4(c) and (d) plot the fitted values of vacuum Rabi splitting and exciton-photon detuning vs the energy of the polariton state. This analysis indicates that the vacuum Rabi splitting is virtually invariant across the whole sample area and has an average value of (116 ± 1.5) meV, with the exciton-photon detuning spanning the range 120 meV, $\delta \in [-240, -120]$ meV. From this, we plot the dependence of $|X_{k_\parallel=0}|^2$ on the energy of the polariton state, as shown in Fig. 4(e).

Using this approach we can also determine the dependence of the measured energy-shift, $\Delta E$, on $|X_{k_\parallel=0}|^2$, shown in Fig. 5. This indicates that the energy-shift of the polaritons on condensation has a sub-linear dependence on $|X_{k_\parallel=0}|^2$; a result that firmly precludes pair-polariton scattering as the underlying mechanism for the observed blueshift and suggests that polariton-exciton scattering is also unlikely; here the former process would result in a quadratic dependence on $|X_{k_\parallel=0}|^2$ and the latter on a linear dependence (see Eq. 1).

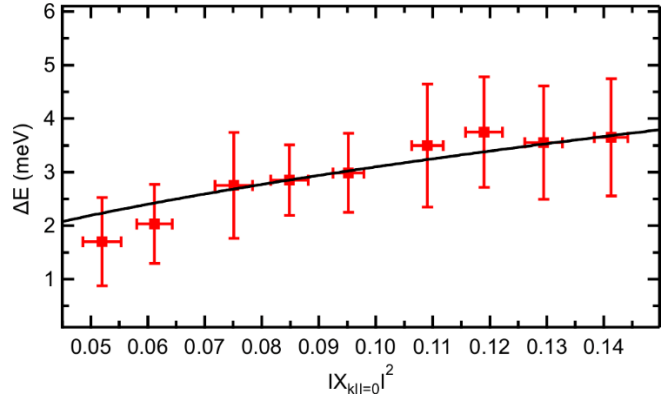

FIG. 5. The blueshift $\Delta E$ versus exciton fraction. The dependence of the blueshift on the exciton fraction is calculated by binning the scattering plot shown in Fig. 4(b) and taking into account the dependence of exciton fraction on the ground polariton state energy (Fig. 4(e)). Black solid line is a fit by sublinear dependence $\Delta E \sim \left(|X_{k_\parallel=0}|^2\right)^{1/2}$

In the absence of pair-polariton interactions and for a constant exciton fraction/detuning (as expressed by Eq. (1)) we expect that polariton-exciton interactions should lead to a linear energy-shift with increasing excitation and thus exciton density. At condensation threshold, stimulated relaxation from the exciton reservoir to the polariton ground-state would lead to a clamping of the exciton density and therefore of the energy-shift. However, to date all non-crystalline semiconductor microcavities undergo a nearly step-wise increase of polariton blueshift at condensation threshold [3–8] as shown in Fig. 2(d), subject to the accuracy of the measured excitation density. Thus, the step-wise dependence precludes polariton-exciton interactions as the driving mechanism for the observed blueshifts; a conclusion that is also corroborated by the sub-linear dependence of the energy-shift on $\left|X_{k_\parallel=0}\right|^2$. Such a conclusion is also consistent with the high degree of localization of Frenkel excitons on a single molecule, as such exciton localization is expected to dramatically weaken Coulomb exchange interactions and suppress interparticle scattering.

In the following, we propose a new mechanism that describes the observed blueshifts, based on a quenching of the Rabi splitting due to the saturation of molecular optical transitions. Owing to the Pauli-blocking principle, excited (i.e. occupied) states cannot be filled twice. Therefore, occupied states do not contribute to the optical absorption at the exciton resonance and effectively reduce the Rabi splitting [4,19] through the relation

$$\hbar\Omega = \hbar\Omega_0 \sqrt{1 - \frac{2(n_x + n_p)}{n_0}} \qquad (2)$$

(see the details in SM, section V). Here, Eq. (2) describes the quenching of the vacuum Rabi splitting, $\hbar\Omega_0$, as a function of the total number of excitations, namely the sum of excitons and polaritons $n_x + n_p$, where $n_0$ is the total number of molecules contributing to the strong coupling. We note here that only a small fraction of excited-state molecules in the intracavity layer couple to the cavity photon, as has been described by Agranovich *et al* [20]. Since the optical pump results in a saturation of the molecular optical transitions that contribute to strong coupling, we expect a partial quenching of the Rabi splitting; an effect that results in a measurable blueshift of the lower polariton mode with increasing excitation density. It is clear however that Eq. (2) by itself does not result in the ubiquitous step-wise increase of the blueshift at condensation threshold, but instead predicts a continuous increase of the blueshift with increasing particle number.

To explain the blueshift dependence, we construct a model that distinguishes between molecules that have a non-zero projection of their optical dipole moment aligned parallel ($N_0^\parallel$) and perpendicular ($N_0^\perp$) to the linear polarization of the excitation laser. We assume that upon non-resonant optical excitation, only parallel-aligned molecules are initially occupied. These molecules constitute an exciton reservoir ($N_x^\parallel$) whose population is then depleted through (i) energetic relaxation to the ground polariton state having the same optical alignment ($N_p^{\parallel}$), (ii) intermolecular energy transfer to perpendicular-aligned molecules as well as to other uncoupled molecules having some out-of-plane projection of the dipole moment $N_0^{\times}$ and (iii) decay via other non-radiative channels ($\gamma_{NR}$). We propose that intermolecular energy transfer from exciton reservoir $N_x^{\parallel}$ populates exciton reservoirs $N_x^{\perp}$ and $N_x^{\times}$, whose populations are in turn depleted through the same energy relaxation channels with the $N_x^{\perp}$ reservoir creating polaritons having an optical alignment that is perpendicular to the excitation laser ($N_p^{\perp}$).

In densely-packed organic films, intermolecular energy transfer is an efficient process that results in the ultrafast depolarization of fluorescence [21,22]. When such films are embedded in a strongly coupled microcavity, intermolecular energy transfer below condensation thresholds is evidenced by a near-zero degree of linear polarization as shown in Fig. 2(e). With increasing excitation density and upon condensation threshold, energy relaxation to the ground polariton state becomes stimulated, resulting in sub-picosecond relaxation times, i.e. stimulated relaxation becomes faster than intermolecular energy transfer. Ipso facto polariton condensation occurs with optical alignment parallel to the excitation laser [3]. The interplay between stimulated relaxation to the ground polariton state and intermolecular energy transfer can qualitatively describe the step-wise increase of the degree of linear polarization at condensation threshold, experimentally observed here in Fig. 2(e). The quenching of intermolecular energy transfer upon condensation threshold, effectively increases the occupation of $N_0^{\parallel}$-molecules, which in turn quenches the corresponding Rabi-splitting, $\hbar\Omega^{\parallel} = \hbar\Omega_0^{\parallel}\sqrt{1 - \frac{2(N_x^{\parallel}+N_p^{\parallel})}{N_0^{\parallel}}}$, and blueshifts the ground polariton state, $N_p^{\parallel}$, by $\Delta E = 1/2 \cdot (E_x + E_c - \sqrt{(E_c - E_x)^2 + (\hbar\Omega^{\parallel})^2}) - E_{LPB}^0$, where $E_x$, $E_c$ are the energies of the bare exciton and cavity modes respectively, and $E_{LPB}^0$ is the energy of ground polariton state in the limit of small excitation numbers (linear regime). The competition between stimulated relaxation to the ground polariton state and intermolecular energy transfer qualitatively predicts the saturation of molecular optical transitions that are optically aligned with the excitation laser, and the concomitant step-wise energy-shift at condensation threshold shown in Fig. 2(d).

To quantitatively describe the experimental dependence of the polariton emission intensity, energy-shift and degree of linear polarization with increasing excitation density, shown in Figs.

2(c-e), we formulate the above model in terms of coupled rate equation (for the details see SM, Section VI):

$$\frac{dN_0^{\parallel,\perp,\times}(t)}{dt} = -P^{\parallel,\perp}(t)N_0^{\parallel,\perp}(t) + N_p^{\parallel,\perp}(t)\gamma_p + N_x^{\parallel,\perp,\times}(t)\gamma_{NR} + N_x^{\parallel,\perp,\times}(t)2\gamma_{xx} - N_x^{\times,\parallel,\perp}(t)\gamma_{xx}$$
$$- N_x^{\perp,\times,\parallel}(t)\gamma_{xx}$$

$$\frac{dN_x^{\parallel,\perp,\times}(t)}{dt} = P^{\parallel,\perp}(t)N_0^{\parallel,\perp}(t) - N_x^{\parallel,\perp}(t)\{N_p^{\parallel,\perp}(t)+1\}\gamma_{xp} - N_x^{\parallel,\perp,\times}(t)2\gamma_{xx} + N_x^{\times,\parallel,\perp}(t)\gamma_{xx}$$
$$+ N_x^{\perp,\times,\parallel}(t)\gamma_{xx} - N_x^{\parallel,\perp,\times}(t)\gamma_{NR}$$

$$\frac{dN_p^{\parallel,\perp}(t)}{dt} = N_x^{\parallel,\perp}(t)\{N_p^{\parallel,\perp}(t)+1\}\gamma_{xp} - N_p^{\parallel,\perp}(t)\gamma_p$$

where $P^{\parallel,\perp}(t)$ is the term corresponding pulsed optical excitation, in the case of linearly polarized pump $P^\perp(t) = 0$, $\gamma_{NR} = 2.5 \cdot 10^8 s^{-1}$ is the nonradiative decay rate of the exciton reservoirs, $\gamma_p = 10^{13} s^{-1}$ is the polariton decay rate, $\gamma_{xx} = 3.33 \cdot 10^{10} s^{-1}$ are decay rates of intermolecular energy transfer between $N_x^\parallel$, $N_x^\perp$ and $N_x^\times$, $\gamma_{xp} = 1.625 \cdot 10^5 s^{-1}$ is the relaxation rate from exciton reservoirs ($N_x^\parallel, N_x^\perp$) towards the ground polariton states ($N_p^\parallel, N_p^\perp$), respectively. We note here that in the energy relaxation from the exciton reservoir to the ground polariton state we have included a stimulation term $N_x^{\parallel,\perp}(t)\{N_p^{\parallel,\perp}(t)+1\}\gamma_{xp}$. The solid lines in Figs. 2(c-e) show the result of the numerical simulations, where we find good quantitative agreement with the experimental observations; for more information, see Section VI in SM. We note here that by switching off intermolecular energy transfer ($\gamma_{xx} = 0$), we obtain a linear dependence of the energy-shift with increasing excitation density that saturates above condensation threshold.

Unlike inorganic semiconductor microcavities that bear Wannier-Mott excitons, interparticle Coulomb exchange interactions are virtually absent in organic microcavities due to the localised nature of Frenkel excitons in molecular semiconductors. In the absence of Coulomb interactions, the origin of blueshifts in organic semiconductor microcavities and in particular the step-wise energy-shift at condensation threshold has remained unclear. We addressed this conundrum by experimentally investigating the contribution of optically induced change of the intracavity non-linear refractive index, gain induced frequency-pulling, and quenching of the Rabi splitting, as well as the role of polariton-exciton and polariton-polariton scattering in the energy-shift of the polariton mode at condensation threshold in strongly coupled molecular dye microcavities. We found that although the quenching of the Rabi splitting leads to a blueshift with increasing excitation density, the linear saturation of molecular transitions cannot explain the step-wise energy-shift at condensation threshold. We attributed the origin of the step-wise blueshift dependence in organic semiconductor microcavities to intermolecular energy transfer. This ultrafast energy migration mechanism is omnipresent in densely-packed organic films and underlies the rapid depolarisation of the emission upon optical excitations. We showed that the step-wise blueshift occurs at condensation threshold, when stimulated relaxation of optically aligned excitons to the ground polariton state exceeds the rate of intermolecular energy transfer. The interplay of intermolecular energy transfer and stimulated exciton relaxation predicts a step-wise increase of the degree of linear polarisation related to the step-wise blueshift at condensation threshold, that is also experimentally observed. We constructed a simple model of the transient dynamics of optically aligned excitons and polaritons that reproduces qualitatively and quantitatively the ubiquitous step-wise blueshift at condensation threshold in non-crystalline organic microcavities.

**Acknowledgments.**

We thank Professor Nikolay Gippius for helpful discussions. AZ and AB acknowledge financial support from Russian Scientific Foundation (RSF) grant No. 18-72-00227. OK and IS acknowledge support from the Government of the Russian Federation (projects 14.Y26.31.0015 and 3.2614.2017/4.6) and ITMO Fellowship Program. We also acknowledge partial funding from the UK EPSRC via Programme Grant 'Hybrid Polaritonics' EP/M025330/1.

**References.**

[1] D. G. Lidzey, D. D. C. Bradley, M. S. Skolnick, T. Virgili, S. Walker, and D. M. Whittaker, Nature **395**, 53 (1998).

[2] S. Kéna-Cohen and S. R. Forrest, Nat. Photonics **4**, 371 (2010).

[3] J. D. Plumhof, T. Stöferle, L. Mai, U. Scherf, and R. F. Mahrt, Nat. Mater. **13**, 247 (2014).

[4] K. S. Daskalakis, S. A. Maier, R. Murray, and S. Kéna-Cohen, Nat. Mater. **13**, 271 (2014).

[5] C. P. Dietrich, A. Steude, L. Tropf, M. Schubert, N. M. Kronenberg, K. Ostermann, S. Höfling, and M. C. Gather, Sci. Adv. **2**, e1600666 (2016).

[6] T. Cookson, K. Georgiou, A. Zasedatelev, R. T. Grant, T. Virgili, M. Cavazzini, F. Galeotti, C. Clark, N. G. Berloff, D. G. Lidzey, and P. G. Lagoudakis, Adv. Opt. Mater. **5**, 1700203 (2017).

[7] D. Sannikov, T. Yagafarov, K. Georgiou, A. Zasedatelev, A. Baranikov, L. Gai, Z. Shen, D. Lidzey, and P. Lagoudakis, Adv. Opt. Mater. **?**, 1900163 (2019).

[8] A. V. Zasedatelev, A. V. Baranikov, D. Urbonas, F. Scafirimuto, U. Scherf, T. Stöferle, R. F. Mahrt, and P. G. Lagoudakis, Nat. Photonics **?**, ? (2019).

[9] G. Lerario, A. Fieramosca, F. Barachati, D. Ballarini, K. S. Daskalakis, L. Dominici, M. De Giorgi, S. A. Maier, G. Gigli, S. Kéna-Cohen, and D. Sanvitto, Nat. Phys. **13**, 837 (2017).

[10] M. Vladimirova, S. Cronenberger, D. Scalbert, K. V. Kavokin, A. Miard, A. Lemaître, J. Bloch, D. Solnyshkov, G. Malpuech, and A. V. Kavokin, Phys. Rev. B **82**, 075301 (2010).

[11] Y. Sun, Y. Yoon, M. Steger, G. Liu, L. N. Pfeiffer, K. West, D. W. Snoke, and K. A. Nelson, Nat. Phys. **13**, 870 (2017).

[12] R. Houdré, C. Weisbuch, R. P. Stanley, U. Oesterle, P. Pellandini, and M. Ilegems, Phys. Rev. Lett. **73**, 2043 (1994).

[13] M. S. Skolnick, T. A. Fisher, and D. M. Whittaker, Semicond. Sci. Technol. **13**, 645 (1998).

[14] C. Ciuti, V. Savona, C. Piermarocchi, A. Quattropani, and P. Schwendimann, Phys. Rev. B **58**, 7926 (1998).

[15] V. M. Agranovich and B. S. Toshich, Sov. Phys. JETP **26**, 104 (1968).

[16] M. Litinskaya, Phys. Rev. B **77**, 155325 (2008).

[17]  M. Sheik-bahae, A. A. Said, and E. W. Van Stryland, Opt. Lett. **14**, 955 (1989).

[18]  A. Kavokin, J. J. Baumberg, G. Malpuech, and F. P. Laussy, Microcavities (Oxford University Press, New York, 2007).

[19]  A. S. Brichkin, S. I. Novikov, A. V Larionov, and V. D. Kulakovskii, Phys. Rev. B **84**, 195301 (2011).

[20]  V. M. Agranovich, M. Litinskaia, and D. G. Lidzey, Phys. Rev. B **67**, 085311 (2003).

[21]  M. H. Chang, M. J. Frampton, H. L. Anderson, and L. M. Herz, Phys. Rev. Lett. **98**, 027402 (2007).

[22]  A. J. Musser, S. K. Rajendran, K. Georgiou, L. Gai, R. T. Grant, Z. Shen, M. Cavazzini, A. Ruseckas, G. A. Turnbull, I. D. W. Samuel, J. Clark, and D. G. Lidzey, J. Mater. Chem. C **5**, 8380 (2017).

[23]  M. Sheik-Bahae, A. A. Said, T.-H. Wei, D. J. Hagan, and E. W. Van Stryland, IEEE J. Quantum Electron. **26**, 760 (1990).

[24]  R. L. Sutherland, D. G. Mclean, and S. Kirkpatrick, Handbook of Nonlinear Optics (Marcel Dekker, New York, 2003).

[25]  M. Arikan, R. T. Pepino, S. Ingvarsson, and I. Shelykh, Superlattices Microstruct. **47**, 139 (2010).

[26]  A. Loudet and K. Burgess, Chem. Rev. **107**, 4891 (2007).

# Supplemental Materials: On the origin of blueshifts in organic polariton condensates

Timur Yagafarov[1], Denis Sannikov[1,2], Anton Zasedatelev[1,3*], Kyriacos Georgiou[4], Anton Baranikov[1], Oleksandr Kyriienko[5,6], Ivan Shelykh[6,7], Lizhi Gai[8], Zhen Shen[8], David Lidzey[4], and Pavlos Lagoudakis[1,3*]

*1. Center of Photonics and Quantum Materials, Skolkovo Institute of Science and Technology, Moscow 121205, Russian Federation.*

*2. Lebedev Physical Institute of the Russian Academy of Sciences, Moscow 119333, Russian Federation.*

*3. Department of Physics and Astronomy, University of Southampton, Southampton SO17 1BJ, United Kingdom.*

*4. Department of Physics and Astronomy, University of Sheffield, Sheffield S3 7RH, United Kingdom.*

*5. NORDITA, KTH Royal Institute of Technology and Stockholm University, Stockholm SE-106 91, Sweden.*

*6. ITMO University, Saint Petersburg 197101, Russian Federation.*

*7. School of Engineering and Natural Sciences, University of Iceland, Reykjavik IS-107, Iceland.*

*8. State Key Laboratory of Coordination and Chemistry, School of Chemistry and Chemical Engineering, Nanjing University, Nanjing 210046, China*

*Authors for correspondence: A.Zasedatelev@skoltech.ru; pavlos.lagoudakis@soton.ac.uk*

### Section I. Sample fabrication

**A.** *BODIPY-G1 neat films preparation*.

Polystyrene (PS) with an average molecular weight of $(M_w) \approx 192\,000$ was dissolved in toluene at 35 mg mL$^{-1}$ following heating at 70 ºC and stirring for 30 min. BODIPY-G1 (1,3,5,7-Tetramethyl-8-phenyl-4,4-difluoroboradiazaindacene) was then dispersed into the PS/toluene inert matrix solution at 10% concentration by mass. The PS/BODIPY-G1 blend was spin-cast onto quartz-coated glass substrates to create neat thin films for absorption, photoluminescence, ASE and nonlinear Z-scan measurements.

**B.** *Microcavity fabrication.*

The studied microcavities utilized double DBR mirrors with the bottom DBR having 10 pairs of alternate layers of $SiO_2/Nb_2O_5$ and the top DBR having 8 pairs. The dielectric materials were deposited with an ion-assisted electron beam ($Nb_2O_5$) and reactive sublimation ($SiO_2$). The bottom mirror was deposited on a quartz-coated glass substrate. A layer of PS/BODIPY-G1 was then spin-cast on top of the bottom DBR. The thickness of the organic active layer was controlled via the rotation speed of the spin coater to achieve $\lambda/2$ microcavities. A thickness gradient, occurring from the spin casting process, permitted different photon-exciton detuning values. For the deposition of the top DBR, the ion gun was kept turned-off for the first few layers of $Nb_2O_5/SiO_2$ to avoid any damage on the organic active layer.

## Section II. Experimental setups

**A.** *Linear Spectroscopy.*

Absorption (Abs) of the bare BODIPY-G1 thin-film was measured using a Fluoromax 4 fluorometer (Horiba) equipped with a Xe-lamp. Photoluminescence (PL) of the film was measured using an Andor Shamrock CCD spectrometer following 473 nm laser diode optical excitation. Fig. S1 shows normalized Abs and PL spectra of the neat 150 nm thin film.

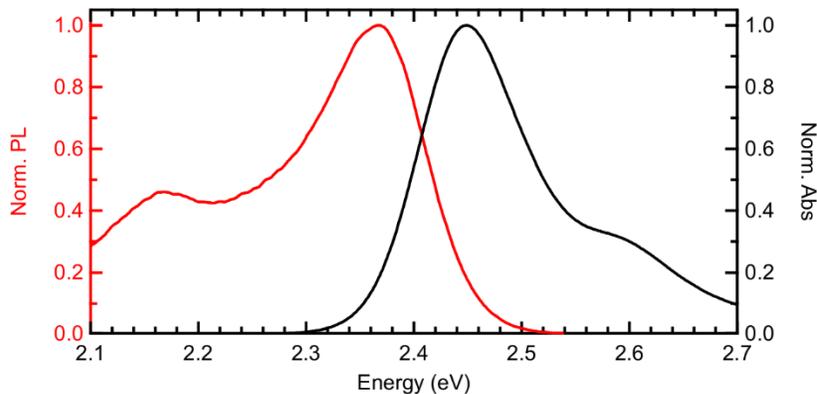

FIG. S1. Normalized absorption (black) and photoluminescence (red) spectra of the bare BODIPY-G1 thin-film.

A fiber-coupled Halogen-Deuterium white light source (Ocean Optics DH-2000) was used for the angular reflectivity measurements. Two motorized optical rails connected to two concentric rotation mounts allowed the illumination of the sample at different angles and the collection of the reflected white light accordingly. An attached fiber bundle at the end of the collection optical rail was used to direct light into an Andor Shamrock charge-coupled device (CCD) spectrometer. The polariton modes were fitted with Lorentzian curves in order to extract the energy of the LPB and UPB at different angles.

**B.** *Single-pulse dispersion imaging*.

We investigated polariton condensate emission using 2 ps single pulses optical excitation from a Ti:Sapphire laser (Coherent Libra-HE) which was frequency doubled through a barium borate crystal providing a wavelength of 400 nm. The pump beam was focused onto a sample by Nikon Plan Fluor 4X microscope objective in 12 µm pump spot size at FWHM. Photoluminescence was collected in transmission configuration using Mitutoyo Plan Apo 20X microscope objective with a numerical aperture of 0.42. To block the residual light from the excitation beam a Semrock LP02-442RU longpass filter was used in the collection path. Filtered photoluminescence from a microcavity was coupled into a 750 mm focal length spectrometer (Princeton Instruments SP2750) equipped with an electron multiplying charge-coupled device camera (Princeton Instruments ProEM-HS 1024 × 1024). 1200 grooves mm$^{-1}$ grating and 20 µm entrance slit were used to achieve a spectral resolution of 30 pm.

**C.** *ASE measurements*.

ASE measurements were performed using a 355 nm pulsed laser with 350 ps pulse width and 1 kHz repetition rate. A 25 mm cylindrical lens was used to focus the beam on the sample creating a stripe excitation profile (1470 µm x 80 µm). ASE emission from the neat 150 nm thin film was detected from the edge of the film, in a direction perpendicular to that of the propagation of the incident pump beam. An Andor Shamrock CCD spectrometer was used to record ASE spectra. The power threshold for ASE was defined at P = 6 mW. Fig. S2 shows ASE spectra recorded in the case of below P < Pth (dashed curve) and above P > Pth (solid curve) amplification threshold.

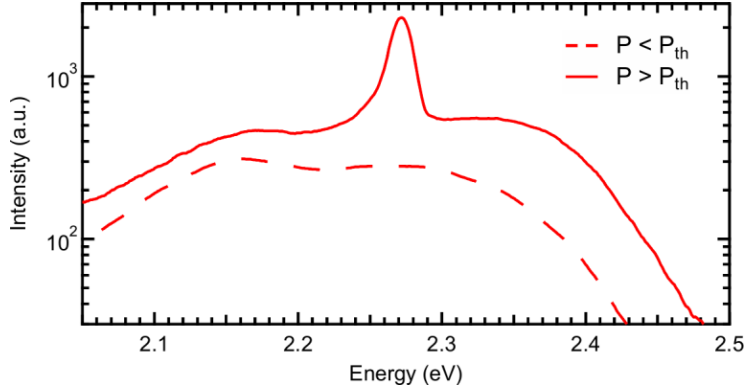

FIG. S2. Amplified spontaneous emission (ASE) of the neat 150 nm thin film below $P < P_{th}$ (dashed curve) and above $P > P_{th}$ (solid curve) amplification threshold.

**D.** *Z-scan measurements*.

We studied two thin films of BODIPY-G1 in polystyrene host matrix and polystyrene film itself, both 600 nm thickness. Measurements were carried out using optical parametric amplifier (Coherent OPerA SOLO) pumped by high energy Ti:sapphire regenerative amplifier system (Coherent Libra-HE) with the central wavelength emission at 545 nm coinciding with the photon energy of the particular polariton condensate realization. Pulse duration and repetition rate were 140 fs and 10 Hz respectively. The beam was tightly focused by 100 mm focal-length lens resulting in 16 µm spot radius. Data acquisition was performed using Si photodetectors (Thorlabs-Det10/M) connected with an oscilloscope (Keysight DSOX3054T). Fig. S3 depicts a sketch of the experimental setup, where PD2 and PD3 are photodetectors recording open and closed aperture signals respectively, PD1 is a reference photodetector. The sample is placed on the motorized translation stage (Thorlabs TravelMax 50 mm driven by Thorlabs 50 mm Trapezoidal Stepper Motor Drive).

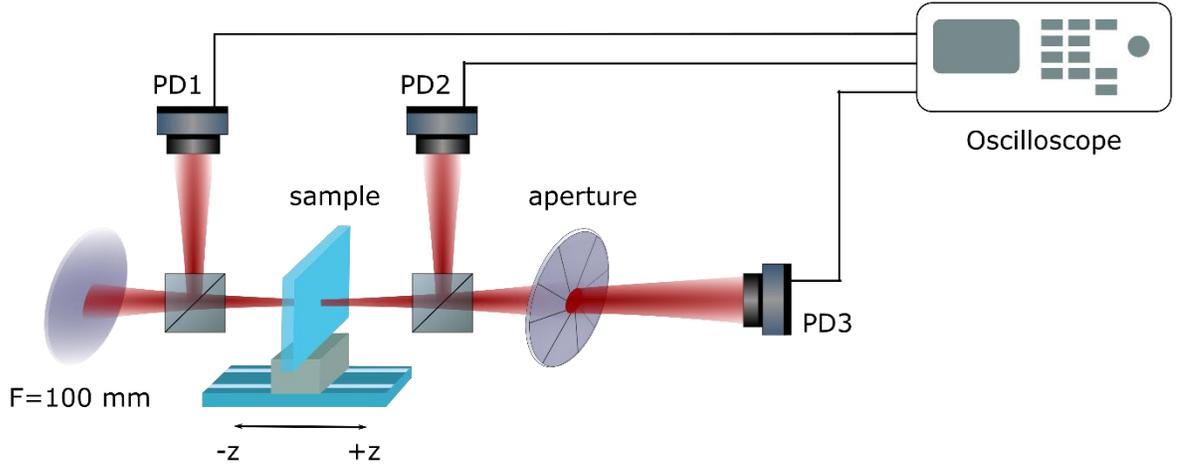

FIG. S3. Schematic of open and closed aperture Z-scan setup. PD1,2,3 are Si photodetectors, the sample is placed on the motorized translation stage moving along z-axis.

## Section III. Data analysis

**A.** *Polariton dispersion relations*

In order to reconstruct polariton dispersion relations we combine angular dependent reflectivity and photoluminescence data. Fig. 1(b) shows excellent agreement of the local spectral minima extracted from the angular dependent reflectivity with *E,k*-distribution of photoluminescence. We extracted vacuum Rabi splitting and exciton-cavity mode detuning by fitting the resulting dispersion relations using a coupled harmonic oscillator model described by the Hamiltonian:

$$\hat{H} = \begin{pmatrix} E_X & \frac{\hbar\Omega_0}{2} \\ \frac{\hbar\Omega_0}{2} & E_C \end{pmatrix}$$

Here, $E_C = \frac{E_C(0)}{\sqrt{1-\sin^2(\alpha)/n_{eff}^2}}$ is an angular dependent cavity photon mode, $n_{eff} = 1.81$ is the effective refractive index of the cavity, $E_C(0)$ is the cut-off energy of the cavity mode (variable parameter), $E_X = 2.446 eV$ is exciton energy eigenstate $S_{1,0}$, $\frac{\hbar\Omega_0}{2}$ is the interaction potential of the exciton-cavity photon mode coupling for the $S_{0,0} \rightarrow S_{1,0}$ singlet-singlet molecular transition

(variable parameter). From the above Hamiltonian one can get standard expressions for upper (UPB) and low (LPB) polariton branches in the following explicit form:

$$E_{UPB,LPB} = \frac{1}{2}\left(E_C + E_X \pm \sqrt{\delta^2 + (\hbar\Omega_0)^2}\right)$$

where $\delta = E_C - E_X$ is the exciton-photon detuning and $\hbar\Omega_0$ is the vacuum Rabi splitting

A polariton wave-function in terms of bare photon and exciton states can be expressed as follows:

$$\psi_{pol} = |X|^2\psi_X + |C|^2\psi_C$$

where $|X|^2$ and $|C|^2$ are Hopfield coefficients related to an excitonic and photonic contributions respectively. Hopfield coefficients can be found from:

$$|C|^2, |X|^2 = \frac{1}{2}\left(1 \pm \frac{\delta}{\sqrt{\delta^2 + (\hbar\Omega_0)^2}}\right)$$

**B.** *Binning*.

The binning procedure for the single realizations of the polariton energy-shift depicted as the scattering plot in Fig. 4(b) have been carried out by Sturge rule. According to the rule a number of bins is $k = 1 + log_2(n)$, where $n$ is the total number of observations. Taking into account the number of experimental realizations equals 378 we get $k = 9.56$, i.e. 9 bins that corresponds to bin width of 14.1 meV in terms of energy. The same approach is relevant to dependence of energy-shift on the exciton fraction. It results in 9 bins of 0.0112648 bin width in terms of $|X_{k_\parallel=0}|^2$.

### Section IV. Extraction of nonlinear susceptibility from Z-scan measurements

To estimate intracavity intensity we recorded a real space image of emission from MC corresponding polariton condensate by using a single pulse imaging technique. Initially, we calibrate the EMCCD camera with a 550 nm 140 fs laser pulse with known single pulse energy generated from Coherent OPerA SOLO. Following that, we measured the pulse energy $W_c$ of outgoing emission from the MC at the polariton condensation ($P_{pump}$~1.4 $P_{th}$). Taking into account the real space profile of the condensate $r = 8\mu m$, the pulse energy $W_c = 0.5 pJ$ and polariton lifetime $\tau = 100 fs$ we assess the outgoing emission intensity as

$$I_0 = \frac{2 \cdot W_c}{\tau \cdot \pi \cdot r^2} \cong 0.5 \cdot 10^7 \, W/cm^2$$

Intracavity field is presented by a standing electromagnetic wave with decaying oscillations:

$$\mathcal{E}_c(t,x) = \mathcal{E}_c(x) \cdot exp\left(\frac{-\omega_c t}{Q}\right) \cdot exp(-i\omega_c t)$$

where $Q$ is the quality factor of the MC, $\omega_c$ is the cavity eigenfrequency and $\mathcal{E}_c(x)$ represents the spatial field distribution. The maximum intracavity intensity is reached at $t = 0$ in the antinode of the standing wave: $I_c = \left|\mathcal{E}_c\left(0, \frac{L_c}{2}\right)\right|^2$, where $L_c = \frac{\lambda_c}{2n}$ is the cavity length.

For the sake of simplicity, we consider a symmetric cavity with two identical DBRs with the same reflectance $R$. Therefore, the outgoing intensity $I_0$ recorded along the one direction of the MC emission is coupled with the intracavity intensity $I_c$ as $I_0 \cong (1-R) \cdot \frac{I_c}{4}$, which is equivalent to $I_c \cong 4 Q/\pi \cdot I_0$. The pre-factor 4 appears from E-field of the standing wave which is a sum of two equivalent but counter propagating waves $\mathcal{E}_c = \mathcal{E}_+ + \mathcal{E}_-$. Taking into account a quality factor of the MC is $Q = \frac{\lambda}{\Delta\lambda} = 350$, one can calculate intracavity intensity $I_c$ is equal to $2.2 \cdot 10^9 \, W/cm^2$

In order to extract imaginary and real parts of nonlinear susceptibility $\chi^{(3)}$ from Z-scan measurements we fitted the data by well-established analytic expressions for normalized transmittance derived from the solution of nonlinear wave equation in the paraxial approximation for a Gaussian beam [23]. We extract imaginary part of nonlinear susceptibility $Im[\chi^{(3)}]$ from open-aperture Z-scan curves by using the following equation:

$$T_{open}(z) = \sum_{m=0}^{\infty} \frac{(-\beta I_0 L_{eff})^m}{\left(1 + \left(\frac{z}{z_R}\right)^2\right)^m (m+1)^{\frac{3}{2}}}$$

where $I_0 = 7.79 * 10^{11} \frac{W}{cm^2}$ – on-axis peak intensity of the excited E-field in focus ($z = 0, r = 0, t = 0$), $L_{eff} = \frac{(1-e^{-\alpha L})}{\alpha}$ – effective thickness of the film, $\alpha = 168.63 cm^{-1}$ – linear absorption coefficient, $L = 600\ nm$ – thickness of the film, $z_R = 2\ mm$ is Rayleigh length of incident Gaussian beam, $\beta$ – nonlinear absorption coefficient, which is the only variable parameter.

The best fit result is shown in Fig. 3(a) [red curve] of the main manuscript. Nonlinear absorption coefficient is equal to $\beta = 3.42 \cdot 10^{-9}$ cm/W according the fitting procedure. The real part of nonlinear susceptibility $Re[\chi^{(3)}]$ is extracted from closed-aperture Z-scan curves using well-known analytic expression below:

$$T_{closed}(z) = 1 + \frac{\frac{8\pi}{\lambda} L_{eff} n_2 I_0 \left(\frac{z}{z_R}\right)}{\left(\left(\frac{z}{z_R}\right)^2 + 9\right)\left(\left(\frac{z}{z_R}\right)^2 + 1\right)}$$

where, $\lambda = 550\ nm$ is the wavelength of the beam, $n_2$ is the nonlinear refractive index which is the only variable parameter. We extract positive nonlinear refractive index $n_2 = 1.89 \cdot 10^{-14}$ cm$^2$/W from the fitting of experimental data as plotted in Fig. 3(b).

Finally we express the nonlinear refractive index $n_2$ and absorption coefficient $\beta$ in terms of real and imaginary part of third order susceptibility [24]:

$$\chi^{(3)} = 2.17 \times 10^{-20} + i \cdot 1.71 \times 10^{-20}\ \left\{\frac{m^2}{V^2}\right\} = 1.55 \times 10^{-12} + i \cdot 1.23 \times 10^{-12}\ \{esu\}$$

### Section V. Microscopic model of the quenching of the Rabi splitting

The origin of the nonlinear optical coupling between a cavity mode and Frenkel excitons comes from the observation that excitons correspond to bosons only in the limit of small excitation numbers. This is easy to see in the case of Frenkel excitons, where each excitation can be described by the Pauli creation (annihilation) operator $\sigma_j^+$ ($\sigma_j^-$), which are two-level raising and lowering operators, acting at the site $j$. The corresponding $\sigma_j^z$ operator then describes the presence of an excitation (akin to density operator). In the limit when the total number of excitations $n_T$ is much less than the total number of the sites $n_0$, one can perform Holstein-Primakoff transformation, collective excitation for the ensemble of two-level systems becomes a boson, and the system consisting of a cavity plus and active organic material can be nicely described by the model of two coupled bosonic modes. However, once the excitation power grows and condition $n_T<<n_0$ does not hold any more, the effects of the saturation of the absorption start to play role and simple model of two coupled oscillators breaks down.

To build the microscopic model corresponding to this regime, we describe a delocalized Frenkel exciton in the momentum-space form using the Fourier transformed operators: $\sigma_k^+ = \frac{1}{n_0}\sum_{j=1}^{n_e} \sigma_j^+ e^{ikr_j}$ where $k$ is an in-plane wavevector for the exciton, and the density operator is defined through commutation relations $[\sigma_k^+; \sigma_{k'}^z] = -2\sigma_{k+k'}^+$, $[\sigma_k^-; \sigma_{k'}^z] = 2\sigma_{k-k'}^-$. The Hamiltonian for Frenkel excitons coupled to light reads

$$\hat{H}_0 = \frac{1}{2}\sum_k E_X(k)\sigma_k^z + \sum_k E_C(k)a_k^+ a_k + \hbar g \sum_k (\sigma_k^+ a_k + \sigma_k^- a_k^+)$$

where the first term describes excitons with dispersion $E_X(k)$ the second term corresponds to the cavity mode with dispersion $E_C(k)$, and the last term describes light-matter coupling term with characteristic constant $g$. The operators $a_k^+(a_k)$ correspond to bosonic creation (annihilation) operators of the quantized electromagnetic cavity mode.

To describe the optical response of the system, we calculate the Green's function of a cavity photon [25] $\langle\langle a_k | a_k^+ \rangle\rangle_t = -i\theta(t)\langle [a_k(t); a_k^+(0)] \rangle$ where index t means that we work in the time domain, and the time dependent operators $a_k(t)$ can be readily obtained from the Heisenberg equations of motion. We then perform the Fourier transform of $\langle\langle a_k | a_k^+ \rangle\rangle_t$ and get the following equation:

$$(\hbar\omega - E_c(k))\langle\langle a_k | a_k^+ \rangle\rangle_\omega = 1 + \hbar g \langle\langle \sigma_k^- | a_k^+ \rangle\rangle_\omega$$

Here, the last term appears due to the light-matter coupling. In a same way the equation of motion for propagator for $\langle\langle \sigma_k^- | a_k^+ \rangle\rangle_\omega$ can be derived, and the system of the equations can be closed by performing Wick-decoupling at the mean-field level which yields

$$\sum_k [(\hbar\omega - E_X(k))(\hbar\omega - E_C(k))\delta_{k,k'} - 2\hbar^2 g^2 \langle \sigma_{k-k'}^z \rangle]\langle\langle a_{k'} | a_{k'}^+ \rangle\rangle_\omega = 1$$

The spectrum can be defined from the poles of the Green function corresponding to the zeroes of the determinant of the system. This gives the following equation for the dispersions of the polariton modes:

$$E_{UPB,LPB}(n_T) = \frac{1}{2}\left(E_C + E_X \pm \sqrt{(E_C - E_X)^2 + \hbar^2 \Omega^2(n_T)}\right)$$

Where the characteristic vacuum Rabi splitting decreases with increase of the total number of excitations in the system due to the effects of the saturable absorption and reads:

$$\hbar\Omega(n_T) = 2\hbar g\sqrt{n_0 - 2n_T} = \hbar\Omega_0\sqrt{1 - \frac{2n_T}{n_0}}$$

where $\Omega_0 = 2g\sqrt{n_0}$ is characteristic vacuum Rabi splitting in the linear regime, $n_T = n_x + n_p$ is the total number of excitation, namely sum of excitons and polaritons. Note, that as expected vacuum Rabi splitting decreases with increase of the number of the excitations and becomes zero when half of the molecules become excited.

## Section VI. Numerical simulations

Numerical simulations are based on the rate equations derived for the molecules with parallel-aligned dipole moment $N_0^{\parallel}$ and with orthogonal optical alignment $N_0^{\perp}, N_0^{\times}$. Fig. S4 shows schematic of molecular transitions and relaxation paths of excited states considered within the model

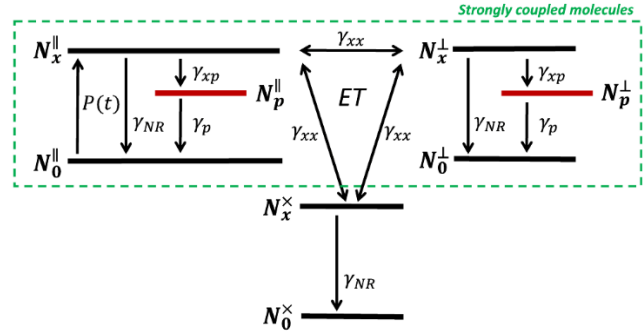

FIG. S4. Schematic of the molecular transitions and relaxation paths of excited states considered within the model.

Explicitly, the rate equations are expressed as follows:

$$\frac{dN_0^{\parallel}(t)}{dt} = -P^{\parallel}(t)N_0^{\parallel}(t) + N_p^{\parallel}(t)\gamma_p + N_x^{\parallel}(t)\gamma_{NR} + N_x^{\parallel}(t)2\gamma_{xx} - N_x^{\times}(t)\gamma_{xx} - N_x^{\perp}(t)\gamma_{xx}$$

$$\frac{dN_0^\perp(t)}{dt} = N_p^\perp(t)\gamma_p + N_x^\perp(t)\gamma_{NR} + N_x^\times(t)2\gamma_{xx} - N_x^\perp(t)\gamma_{xx} - N_x^\times(t)\gamma_{xx}$$

$$\frac{dN_0^\times(t)}{dt} = N_x^\times(t)\gamma_{NR} + N_x^\times(t)2\gamma_{xx} - N_x^\perp(t)\gamma_{xx} - N_x^\parallel(t)\gamma_{xx}$$

$$\frac{dN_x^\parallel(t)}{dt} = P^\parallel(t)N_0^\parallel(t) - N_x^\parallel(t)\{N_p^\parallel(t) + 1\}\gamma_{xp} - N_x^\parallel(t)2\gamma_{xx} + N_x^\times(t)\gamma_{xx} + N_x^\perp(t)\gamma_{xx}$$
$$- N_x^\parallel(t)\gamma_{NR}$$

$$\frac{dN_x^\perp(t)}{dt} = -N_x^\perp(t)\{N_p^\perp(t) + 1\}\gamma_{xp} - N_x^\perp(t)2\gamma_{xx} + N_x^\parallel(t)\gamma_{xx} + N_x^\times(t)\gamma_{xx} - N_x^\perp(t)\gamma_{NR}$$

$$\frac{dN_x^\times(t)}{dt} = -N_x^\times(t)2\gamma_{xx} + N_x^\perp(t)\gamma_{xx} + N_x^\parallel(t)\gamma_{xx} - N_x^\times(t)\gamma_{NR}$$

$$\frac{dN_p^\parallel(t)}{dt} = N_x^\parallel(t)\{N_p^\parallel(t) + 1\}\gamma_{xp} - N_p^\parallel(t)\gamma_p$$

$$\frac{dN_p^\perp(t)}{dt} = N_x^\perp(t)\{N_p^\perp(t) + 1\}\gamma_{xp} - N_p^\perp(t)\gamma_p$$

The pump term $P(t)$ corresponds to the linearly polarized 2 ps pulsed non-resonant optical excitation with Gaussian temporal profile. Effective non-radiative decay rate of the exciton reservoirs $\gamma_{NR} = 2.5 \cdot 10^8 s^{-1}$ is calculated from known values of quantum yield and total lifetime of excited states according to [22]. The rates of intermolecular energy transfer corresponding depolarization processes are taken from time-resolved polarization anisotropy measurements. According to the study [22], polarization decay time for the material is less than 50 ps (limited by the setup time-resolution), thus we use the reasonable value of $\gamma_{xx} = 3.33 \cdot 10^{10} s^{-1}$. Polariton lifetime has been calculated from the full width at half maximum of the emission linewidth below the threshold (1.5 nm) as 100 fs, so that the decay rate is equal $\gamma_p = 10^{13} s^{-1}$ regardless optical alignment since the ground polariton state is degenerate for TE/TM modes at $k \sim 0$.

We estimated total dye density $N_{tot}$ in the cavity as $1.3 \cdot 10^{20}$ molecules per cm³ using Beer-Lambert law taking into account experimental values of the film thickness L = 170 nm, absorbance at 507 nm $A = 0.34$, molar extinction ratio $\varepsilon = 95500$ [26] and molecular weight of the dye M = 384,18. However, the only small fraction of molecules randomly distributed in the cavity undergoes strong light-matter interaction. We introduce a coupling coefficient $F_c$ defining the density of coupled molecules as $N_0^{\|} = N_0^{\perp} = F_c \cdot N_{tot}$. The coupling coefficient $F_c$ as well as exciton-to-polariton relaxation rates $\gamma_{xp}$ are the variable parameters. We impose an experimental criterion for them satisfying polariton condensation that appears at the exciton density in 10% of the total dye density $N_{tot}$. We assume that both exciton reservoirs exhibit the same exciton-to-polariton decay rate $\gamma_{xp}$ which is reasonable approximation since the ground polariton state is fully degenerate in terms of polarization and the molecules are randomly distributed over the cavity volume. By applying the following boundary conditions:

$$N_0^{\|}(t=0) = N_0^{\perp}(t=0) = N_0^{\times}(t=0) = F_c \cdot N_{tot}$$

$$N_x^{\|}(t=0) = N_x^{\perp}(t=0) = N_x^{\times}(t=0) = N_p^{\|}(t=0) = N_p^{\perp}(t=0) = 0$$

we simulated the total time integrated PL as a function of pump power:

$$PL = \int_0^T \{N_p^{\|}(t) + N_p^{\perp}(t)\} \, dt$$

where an upper bound of the numerical integration $T = 3\tau$ ($\tau = 4$ ns effective lifetime of excited states of the dye molecule [22] is used to ensure reliable results as for below and above the condensation threshold.

We fit the experimental dependence of the integrated PL on the pump power by using the model with $F_c$ and $\gamma_{xp}$ are being variable parameters. Fig. 2(c) [black] of the main manuscript shows the best fit result, where $F_c = 0.073$ and $\gamma_{xp} = 1.625 \cdot 10^5 \, s^{-1}$.

The energy of the ground polariton state has been calculated numerically using time-dependent exciton and polariton densities $N_x^{\|,\perp}(t), N_p^{\|,\perp}(t)$ from the rate equations.

$$E_{LPB}^{\|,\perp}(t) = 1/2 \cdot (E_x + E_c - \sqrt{(E_c - E_x)^2 + \{\hbar\Omega^{\|,\perp}(t)\}^2})$$

where $E_x = 2.446 \, eV$ and $E_c = 2.254 \, eV$ are the energies of the bare exciton and cavity modes respectively, $\hbar\Omega^{\|,\perp}(t) = \hbar\Omega_0^{\|,\perp} \sqrt{1 - 2\frac{N_x^{\|,\perp}(t) + N_p^{\|,\perp}(t)}{N_0^{\|,\perp}}}$ is the quenched vacuum Rabi splitting, $\hbar\Omega_0^{\|,\perp} = 116 \, meV$ is the average vacuum Rabi splitting measured across the detuning range at the limit of zero occupancy.

To compare the result of modelling with experimentally observed values we calculated time-integrated blueshift.

$$E_{LPB}^{\|,\perp} = \frac{\int_0^T E_{LPB}^{\|,\perp}(t) \cdot N_p^{\|,\perp}(t)\, dt}{\int_0^T N_p^{\|,\perp}(t)\, dt}$$

$$E_{LPB}^T = E_{LPB}^{\|} \cdot \frac{1}{PL}\int_0^T N_p^{\|}(t)\, dt + E_{LPB}^{\perp} \cdot \frac{1}{PL}\int_0^T N_p^{\perp}(t)\, dt$$

Fig. 2(d) of the main manuscript demonstrates quantitative agreement of the theory with experimental data. It is worth noting in Fig. 2(d) we plot relative blueshift defined as $\Delta E = E_{LPB}|_P - E_{LPB}|_{P=0.8P_{th}}$ as extracted from experimental data.

To demonstrate the role of the ET in the polariton dynamics we purposely forbid the processes of energy migration between exciton reservoirs assuming $\gamma_{xx} = 0$. Fig. S5 shows the energy of the ground polariton state predicted by the model with (red solid curve) and without intermolecular energy transfer (black dashed curve).

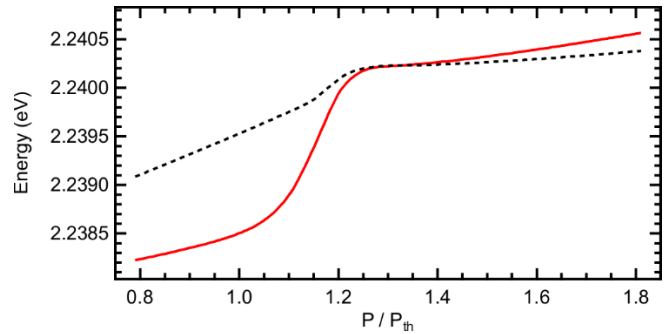

FIG. S5. The ground polariton state energy versus relative pump power (P/P$_{th}$) simulated numerically using the model with (red solid curve) and without intermolecular energy transfer (black dashed curve).

In order to model degree of linear polarization (DLP) we used time-dependent parallel and perpendicular aligned polariton densities $N_p^{\|}(t)$ and $N_p^{\perp}(t)$ respectively. Since we use horizontally polarized optical excitation time-integrated DLP has been calculated in accordance with the following expression:

$$DLP = \frac{\int_0^T \{N_p^{\parallel}(t) - N_p^{\perp}(t)\} dt}{\int_0^T \{N_p^{\parallel}(t) + N_p^{\perp}(t)\} dt}$$

Fig. 2(e) of the main manuscript shows the result of numerical simulation of DLP versus pump power. Note, the condensate above the threshold follows horizontal polarization as initially induced by the pump beam.